\documentclass[twocolumn,aps,prl,showpacs,groupedaddress,amsfonts,amssymb,amsmath]{revtex4-2}
\usepackage{xcolor}
\usepackage{graphicx}
\usepackage{epstopdf}
\usepackage[colorlinks=true, letterpaper=true, pdfstartview=FitV, linkcolor=blue, citecolor=blue, urlcolor=blue]{hyperref}

\begin{document}

\title{Topological photonic alloy}
\author{Tiantao Qu,$^{1,*}$ Mudi Wang,$^{2,*}$ Xiaoyu Cheng,$^{3}$ Xiaohan Cui,$^{2}$ Ruo-Yang Zhang,$^{2}$ Zhao-Qing Zhang,$^{2}$  Lei Zhang,$^{3,4,\dagger}$ Jun Chen,$^{1,4,\ddagger}$ C. T. Chan$^{2,\S}$}
\address{$^1$State Key Laboratory of Quantum Optics and Quantum Optics Devices, Institute of Theoretical Physics, Shanxi University, Taiyuan 030006, China\\
$^2$Department of Physics, The Hong Kong University of Science and Technology, Clear Water Bay, Hong Kong 999077, China\\
$^3$State Key Laboratory of Quantum Optics and Quantum Optics Devices, Institute of Laser Spectroscopy, Shanxi University, Taiyuan 030006, China\\
$^4$Collaborative Innovation Center of Extreme Optics, Shanxi University, Taiyuan 030006, China}

\begin{abstract}
We present the new concept of photonic alloy as a non-periodic topological material. By mixing non-magnetized and magnetized rods in a non-periodic 2D photonic crystal configuration, we realized photonic alloys in the microwave regime. Our experimental findings reveal that the photonic alloy sustains non-reciprocal chiral edge states (CESs) even at very low concentration of magnetized rods. The non-trivial topology and the associated edge states of these non-periodic systems can be characterized by the winding of the reflection phase. 
Our results indicate that the threshold concentrations for the investigated system within the first non-trivial band gap to exhibit topological behavior approach zero in the thermodynamic limit for substitutional alloys, while the threshold remains non-zero for interstitial alloys. At low concentration, the system exhibits an inhomogeneous structure characterized by isolated patches of non-percolating magnetic domains that are spaced far apart within a topologically trivial photonic crystal. Surprisingly, the system manifests CESs despite a local breakdown of time-reversal symmetry rather than a global one. Photonic alloys represent a new category of disordered topological materials, offering exciting opportunities for exploring topological materials with adjustable gaps.
\end{abstract}
\maketitle
Topological materials have become a prominent area of research in various fields of physics \cite{Hasan1,Zhang2,Kane3,Zhang4,Ozawa5,Lu6,Kim7,Haldane8,Wang9,Hu10,Barik11,Shalaev12,Yang13,Wang14,Pai15,Khanikaev16,Zhaoju17,He18,Jiuyang19,Ding20,Mudi21,Guancong22,Haoran23}. These materials are defined by their bulk non-trivial topological invariants, which lead to phenomena like topological edge states that are robust against local perturbations. Originally studied in condensed matter systems \cite{Thouless24}, topological materials have since been observed in photonic \cite{Ozawa5,Lu6,Kim7,Haldane8,Wang9,Hu10,Barik11,Shalaev12,Yang13,Wang14} and acoustic platforms  \cite{Pai15,Khanikaev16,Zhaoju17,He18,Jiuyang19,Ding20,Mudi21,Guancong22,Haoran23}. It has also been discovered that a periodic crystalline structure is not necessary for bulk topological invariants to exist \cite{Mitchell25,Zhe26,Agarwala27,Yang28,Wang29,Ivaki30,Agarwala31} and photonic systems have proven to be an excellent platform for realizing non-periodic topological materials, such as photonic quasicrystals \cite{Kraus32,Bandres33}, photonic Anderson insulators \cite{Stutzer34,Liu35,Cui36}, and photonic amorphous materials \cite{Zhou37,Bing38}.

While disorder and randomness are often seen as detrimental, disorder can induce non-trivial topology, as demonstrated by the concept of topological Anderson insulator \cite{Stutzer34,Liu35,Cui36,Li39,Groth40,Meier41}. In the field of material science, it is well established that randomness can also give rise to extraordinary and advantageous characteristics in materials through another form known as random alloy. Alloys are a group of non-periodic materials that are composed of a mixture of different chemical elements, in which the random distribution of elements can result in unique physical properties that differ from those of their constituents \cite{Mott42,George43,Hart44}. However, the concept of “alloy” in photonic systems with topological properties has remained unexplored. 
	
\begin{figure}
	\centering
	\includegraphics[width=0.45\textwidth]{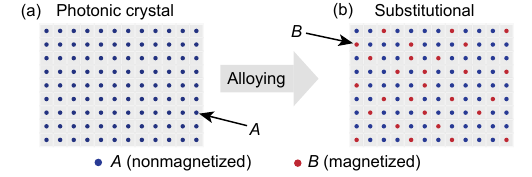}
	\caption{(a) Schematic diagram of photonic crystals with non-magnetized (A) YIG rods. (b) Substitutional alloy with a random mixture of A and magnetized (B) YIG rods.}\label{Fig1}
\end{figure}

In this letter, we introduce a new concept of topological alloys into photonics by demonstrating that non-trivial topology can emerge in disordered 2D photonic crystals composed of random mixtures of non-magnetized (A) and magnetized (B) yttrium iron garnet (YIG) rods in the form of substitutional and interstitial alloys. Remarkably, in these random photonic alloys, chiral edge states (CESs) can emerge even when the concentration $x$ of magnetized rods is very low. 
Instead of applying magnetism everywhere in the crystal to achieve non-trivial topology, we can accomplish the same by just applying magnetism to a few randomly chosen sites. Moreover, we will demonstrate below that there exists a substitutional photonic alloy configuration where the density of these randomly selected positions can approach zero. We experimentally confirm the existence of CESs through non-reciprocal transmission and edge-state distribution. The non-trivial topology of CESs is confirmed through a topological invariant which is defined by the appearance of a 2$\pi$\ reflection phase change in the scattered wave, computed via full-wave simulation. 

\begin{figure}
	\centering
	\includegraphics[width=0.45\textwidth]{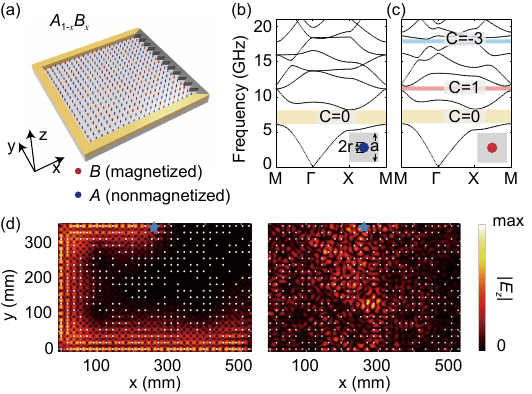}
	\caption{(a) Schematic diagram of the substitutional photonic alloy (${\rm A}_{0.6}{\rm B}_{0.4}$). (b)-(c) The TM bulk band structures of A/B-type photonic crystal, respectively. The lower inset illustrates the lattice geometry. (d) Simulated electric field distribution $|E_z|$  at 11.15 GHz for photonic alloy ${\rm A}_{0.6}{\rm B}_{0.4}$ (left panel) and that with A rods replaced by air (right panel). A-type rods’ positions are marked with white circles and squares. Black circles mark the B-type rods. The line source is denoted with the blue star.}\label{Fig2}
\end{figure}

In an ordinary photonic crystal (crystal A) comprising an array of non-magnetic dielectric rods, as depicted in Fig. \ref{Fig1}(a), topological chiral edge states are absent. In this study, photonic alloys are achieved by introducing some magnetic rods into crystal A. We mainly explore the substitutional random photonic alloys. Figure \ref{Fig1}(b) shows a typical substitutional (${\rm A}_{1-x}{\rm B}_x$) photonic alloy, with the doping concentration $x$ defined as $x = N_{\rm B}/(N_{\rm A}+N_{\rm B})$, with $N_{\rm A}$ and $N_{\rm B}$ representing the number of A-type (non-magnetic) and B-type (magnetic) rods, respectively. To experimentally demonstrate the existence of CESs in the photonic alloys, we consider a rectangular array ($20a \times 30a$), where $a = 18$ mm is the lattice constant, composed of YIG cylindrical rods (4.0 mm diameter, 10 mm height) containing both A-type and B-type rods, as shown in Fig. \ref{Fig2}(a). The B-type YIG rods are individually magnetized using permanent magnets embedded under the rods.

We calculate the transverse magnetic (TM) band structure and the corresponding Chern numbers in the band gaps for both pure A and B photonic crystals, which represent the two limiting configurations of the ${\rm A}_{1-x}{\rm B}_x$ substitutional photonic alloy with $x=0$ and 1 \cite{Skirlo45,Skirlo46}. As expected, the band structure shown in Fig. \ref{Fig2}(b) for the non-magnetic crystal exhibits trivial topology characterized by the zero Chern numbers in all the bands. When $x=1$, all the rods are magnetic and the band structure exhibits nontrivial band topology, as shown in Fig. \ref{Fig2}(c). Specifically, a non-trivial bandgap between the second and third bands emerges, with a Chern number of 1. Our focus lies on the frequency range within this bandgap, as we investigate the topology of the photonic alloys for doping concentrations ranging $0 < x < 1$. The full-wave simulated field pattern under the excitation of a point source at frequency 11.15 GHz is shown in Fig. \ref{Fig2}(d). Here we have enclosed the photonic alloy with metal cladding on three sides as illustrated in Fig. \ref{Fig2}(a). Remarkably, the numerical results indicate the presence of CES bound between the photonic alloy and metal cladding, as shown in the left panel of Fig. \ref{Fig2}(d). This CES propagates unidirectionally in an anti-clockwise direction and can wrap around the corners, indicating that randomly substituting some of the rods in A-type photonic crystal by magnetized rods opens up a topological gap which sustains CESs.

\begin{figure}
	\centering
	\includegraphics[width=0.48\textwidth]{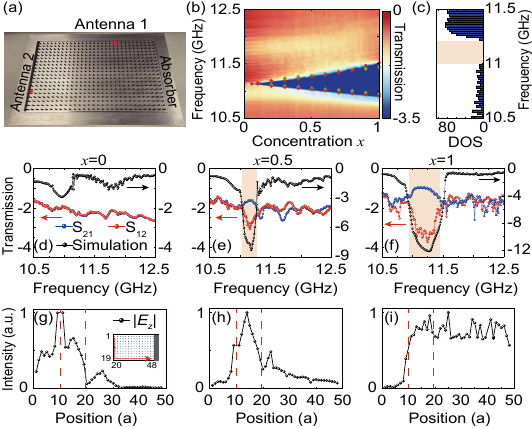}
	\caption{(a) Photo of an experimental sample with a rectangular lattice geometry (top metal plate removed). Two antennas marked with red stars are placed on the top and left edges. (b) Simulated transmission spectrum versus $x$. The red circles represent the topological gap calculated from the reflection phase winding. The green diamonds denote experimental measured non-reciprocal transmission region. (c) Simulated DOS of photonic alloy ($x=0.5$) using 20 configurations. (d)-(f) Experimentally measured edge transmission (solid lines with blue and red dots, left axis) and simulated transmission (solid line with black dots, right axis) when $x=0, 0.5, 1$, respectively. (g)-(i) Experimentally measured edge field distributions when $x=0, 0.5, 1$ at 11.20 GHz, respectively. The first/second red vertical dash lines at position $\#10/20$ indicate the position of the light source and the bottom left corner of the photonic alloy as indicated in the inset of panel (g).}\label{Fig3}
\end{figure}

We compare the system with another disordered configuration, namely replacing the A-type rods with air in the same positions. As shown in the right panel of Fig. \ref{Fig2}(d), the $ {\rm Air}_{0.6}{\rm B}_{0.4}$ system does not support CESs, highlighting the uniqueness and topological non-trivial nature of our proposed photonic alloy. In Fig. S2 of Supplemental Material (SM), we also present the CES distribution for a whole range of filling ratios \cite{supp}. Our observation is that the CES can be sustained in random substitutional photonic alloys and generated robustly even at a very low doping concentration $x$. We found that the presence of a topological gap in the system is essential to sustain CES. Notably, the A-type crystal carries a band touching point between the $2^{\rm nd}$ and the $3^{\rm rd}$ band in Fig. \ref{Fig2}(b), and this degeneracy is broken to open a non-trivial bandgap of the magnetic B-type photonic crystal. For the case of  ${\rm A}_{1-x}{\rm B}_x$, once a small fraction of magnetic rods are randomly introduced into the system, a tiny topological gap will be opened at the band touching point, leading to the emergence of CESs. We will discuss the gap opening later. In contrast, for the case of ${\rm Air}_{1-x}{\rm B}_x$, the percolating air voids support propagating states within the frequency range of the non-trivial bandgap of magnetic B-type photonic crystal. The presence of such states hinders the formation of topological gaps in ${\rm Air}_{1-x}{\rm B}_x$ system required for the formation of CESs. Additional information on ${\rm Air}_{1-x}{\rm B}_x$ system is presented in the SM \cite{supp}. The photonic alloy system exhibits order at the extremes of $x=0$ and $x=1$ while displaying disorder in the intermediate range. We anticipate that the alloy will acquire topological properties once $x$ surpasses a specific threshold value. When the non-magnetic unit cells contain air rods, the system remains non-topological until the magnetic rods form robust interconnected networks at high values of $x$. However, if the dielectric rods are used, the system becomes topological even for extremely small values of $x$. 

To verify our theoretical results, we study experimentally both the phase diagram of the topological gap and the distribution of CESs. The experimental sample is shown in Fig. \ref{Fig3}(a) and we put two antennas near the upper and left metal cladding edges to measure the transmission $S_{ij}$, as indicated by two red stars. Three sides of the array are covered with non-magnetic metal cladding, while the right side is covered with microwave absorbers. More details on materials and the experimental setup can be found in Methods \cite{supp}. At $x=0$, the photonic alloy system is an ordinary non-magnetic photonic crystal and we expect $S_{12}=S_{21}$ due to reciprocity. This observation is evident in Fig. \ref{Fig3}(d), where the red dots ($S_{12}$) and blue dots ($S_{21}$) on the left axis coincide with each other. However, when $x=0.5$, the random substitution of half of the rods by magnetized rods leads to the opening of a gap to host the CESs. Indeed, Fig. \ref{Fig3}(e) clearly shows a frequency region [11.05, 11.28] GHz in which a sharp drop of wave transport occurs in one direction $S_{12}$, which is much smaller than $S_{21}$. This gives evidence to the existence of non-reciprocal CESs. To further verify that the non-reciprocal region indeed lies in a gap, we numerically studied the average transmission of the system by placing a line source at the middle of left side of the system. Under a periodical boundary condition in the vertical direction, we evaluated the energy exiting the right boundary denoted as $E_{\rm out}$ and the summation of energy leaving both the left and right boundaries denoted as $E_{\rm tot}$. The bulk transmission is obtained through $\langle T\rangle=\langle {\rm log}_{10}(E_{\rm out}/E_{\rm tot})\rangle$ \cite{Zhou37,Skirlo46}. The black circles in Fig. \ref{Fig3}(e) shows the result of $\langle T\rangle$ obtained by averaging over 20 configurations at each frequency. The presence of a gap in which $\langle T\rangle$ drops suddenly is clearly seen. Its location coincides well with the frequency window of CESs measured experimentally. Good agreements have also been found in other concentrations in SM Fig. S4 \cite{supp}. In Figs. \ref{Fig3}(d) and \ref{Fig3}(f), we show the results for the special cases of $x=0$ and 1, respectively.

It is important to note that the sudden drop in $\langle T\rangle$ represents a gap in the density of states (DOS). To illustrate this, we conducted a numerical simulation of the DOS for a photonic alloy (${\rm A}_{0.5}{\rm B}_{0.5}$) with 20 configurations, as shown in Fig. \ref{Fig3}(c). We observed a clean DOS gap located in the frequency range of [10.99, 11.21] GHz, which corresponds to the sudden drop in transmission as shown in Fig. \ref{Fig3}(b). In SM Fig. S5, we show that the DOS gap aligns well with the transmission gap for all concentrations studied \cite{supp}.

To further investigate the behaviour of CESs in the substitutional photonic alloys, we measured the $|E_z|$ field distributions experimentally at a frequency of 11.20 GHz on the edge sites. The measured field intensity at various positions along the sample's edge is presented in Figs. \ref{Fig3}(g)-\ref{Fig3}(i). The locations of these measured positions are indicated by the inset in Fig. \ref{Fig3}(g). We excite the TM wave using a source antenna placed at the point labeled as position $\#10$. Figure \ref{Fig3}(g) illustrates that the measured field intensity exhibits a near-symmetrical pattern on both the left and right sides of the source antenna. This shows that the TM wave can propagate equally well in both directions in the A-type photonic crystal as expected. However, in the photonic alloy system, as shown in Fig. \ref{Fig3}(h), the distribution of field intensity becomes significantly asymmetrical around the source with right-hand side (position \#$>10$) being much stronger, indicating directional propagation which gives evidence to the presence of CESs. In Fig. \ref{Fig3}(i), the unidirectional CES is evident when $x=1$. The measured $|E_z|$ field distributions for other concentrations are presented in SM Fig. S6 \cite{supp}.

To see the entire phase diagram of the topological gap, in Fig. \ref{Fig3}(b), we plot the experimentally observed frequency window of CESs at each concentration by two green diamonds. The color map in Fig. \ref{Fig3}(b) shows the numerical results of  $\langle T\rangle$ as a function of concentration and frequency. It is seen that the positions of green diamonds fall on the boundaries of the dark region for all the concentrations studied experimentally. It is also interesting to see that the size of topological gap decreases with decreasing $x$ and vanishes at some threshold concentration $x_{\rm th}\simeq 0.1$ with a frequency close to the band touching point of the $2^{\rm nd}$ and $3^{\rm rd}$ band in the A-type photonic crystal. The finite value of $x_{\rm th}$ found here may be due to the finite size effect. If so, increasing the sample size should be able to push $x_{\rm th}$ to an even smaller value. It is also possible that $x_{\rm th}$ may eventually reduce to zero in the thermodynamic limit.

\begin{figure}
	\centering
	\includegraphics[width=0.45\textwidth]{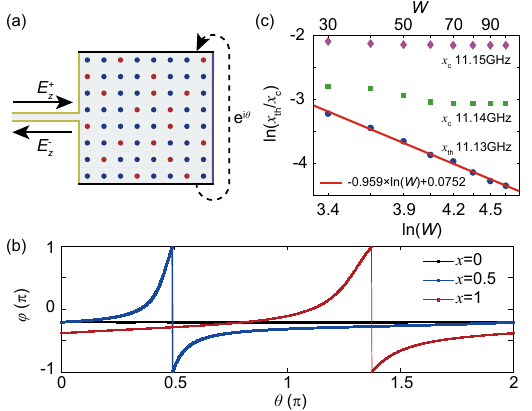}
	\caption{(a) Schematic for retrieving the topological signature of CES from the reflection phase connected to a square photonic alloy with linear size $W$ with a twisted boundary condition $\Psi(y=W)=\Psi(y=0)e^{i\theta}$ imposed to the vertical boundaries. The left side of the photonic alloys is connected with an air lead bounded by PMC (yellow), and the right boundary (purple) is set as scattering boundary condition. (b) Reflection phase winding for the substitutional photonic alloy systems measured at the experiment when $x=0, 0.5, 1$ at 11.20 GHz. (c) Sample size dependence of the threshold concentration $x_{\rm th}$ and critical concentration $x_{\rm c}$. Each data point is computed with 10 configurations.}\label{Fig4}
\end{figure}

To investigate this possibility, we compute the winding of the reflection phases $\varphi$ of a large supercell \cite{Cui36}, which is a rigorous yet computationally efficient method to characterize a topological state in random media. To accomplish this, we connect a waveguide, which supports a single TM mode at the relevant frequency range, to the left side of the photonic alloy. The twisting boundary condition with a twisting angle $\theta$ is imposed along the $y$-direction in Fig. \ref{Fig4}(a). The reflection phases $\varphi$ are calculated as the phase difference between the incident and the reflected waveguide modes under the twisted boundary condition (section 7 in the SM \cite{supp}). In Fig. \ref{Fig4}(b), we show the computed reflection phases for the experimental setup at $f=11.20$ GHz. There is no phase winding for A-type photonic crystal. We observed that the reflection phases $\varphi$ wind 2$\pi$ when the angle $\theta$ is varied by 2$\pi$ for B-type photonic crystal. The winding of one cycle (2$\pi$) is consistent with the existence of one CES in the gap. We then observe that the reflection phases $\varphi$ also wind 2$\pi$ in the ${\rm A}_{0.5}{\rm B}_{0.5}$ photonic alloy. This supports the notion that the presence of substitutional disorder can lead to the emergence of topologically non-trivial states.

We identify the topological gap at each given concentration $x$ by studying the winding of the reflection phase as a function of frequency. A region of quantized non-trivial winding is found between two frequencies for each value of $x$. These two frequencies are marked by two red circles in Fig. \ref{Fig3}(b). We found remarkable agreements between the theory (red circles) and experiments (green diamonds) for all concentrations studied experimentally. It is also noticed that the gap closes at some non-zero threshold concentration $x_{\rm th}$ with the frequency close to the touching frequency of A-type photonic crystal, i.e., $f_t=11.13$ GHz.  To see whether threshold doping concentration is zero in the thermodynamic limit, we study the sample size dependence of $x_{\rm th}$. We fix the frequency at $f_t$ and study $\varphi$ using different sample sizes. For simplicity, we choose square samples of size $\left(W\times W\right)$ and set $a =1$. The results of ${\rm ln}\left(x_{\rm th}\right)$ at each ${\rm ln}\left(W\right)$ are shown by solid blue dots in Fig. \ref{Fig4}(c). A linear behavior in the $\rm ln-ln$ plot implies a power law behavior between the threshold concentration $x_{\rm th}$ (at which 2$\pi$ winding is observed) and the sample size $W$, i.e., $x_{\rm th}\propto W^{-s}$ with an exponent $s=0.959$. Therefore, we predict that the threshold doping concentration of substitutional photonic alloy indeed approaches zero in the thermodynamic limit. This intriguing finding implies that for a large enough sample, even a minute concentration of magnetic rods has the potential to realize topological photonic alloys. 

We have also studied the size dependence of the critical concentration $x_{\rm c}$ of topological transition at frequencies different from the band touching point. The results of $f=11.14, 11.15$ GHz are also shown in Fig. \ref{Fig4}(c). It is found that for those frequencies, $x_{\rm c}$ decreases with $W$ but becomes saturated when $W$ is sufficiently large. In the SM (Fig. S8), we numerically studied the upper and lower phase boundaries of the topological gap, which are well fitted by the analytic relations $f-f_t=0.29x^2+0.11x$ and $f-f_t=-0.26x$, for the upper and lower boundaries of the gap, respectively. Finally, we found that a generalized Haldane model with flux randomly introduced in certain unit cells can mimic the zero threshold phenomenon (section 9 in the SM \cite{supp}). Section 10 of the SM shows that the threshold concentration needed to sustain non-trivial topology depends on the alloy configuration \cite{supp}. For example, an interstitial photonic alloy requires a low, but non-zero, threshold concentration to sustain non-trivial topology.

In conclusion, we realized topological photonic alloys consist of a random mixture of non-magnetized and magnetized rods in the microwave regime. The study demonstrates the emergence of topologically nontrivial gap in these systems, confirmed by existence of robust CESs and the reflection phase winding as a topological signature. The idea has been experimentally validated through measurements of transmission and edge state distribution in substitutional photonic alloys. 
Additionally, our numerical results strongly suggest that the threshold concentrations for the investigated system within the first non-trivial band gap to exhibit topological behavior approach zero in the thermodynamic limit for substitutional alloys.
This implies that the CESs in substitutional photonic alloy can emerge at extremely low doping concentrations without requiring order. Differing from the conventional approach of constructing topological crystals, we showed that alloy-type configurations can sustain non-trivial topology when a very low concentration of topological domains is embedded in non-topological domains. It's expected that the topological photonic alloy can be realized in honeycomb and Kagome lattices (section 11 in the SM \cite{supp}). Our work not only offers a new platform and paves the way for researchers delving into the realm of topological properties in disordered systems, but also show that CESs can be created without the need to break time reversal everywhere inside the crystal and the ideas can be applied to other systems such as acoustic systems. 

This work is supported by the National Key R$\&$D Program of China under Grant No. 2022YFA1404003, the National Natural Science Foundation of China Grant No. 12074230, 12174231, the Research Grants Council of Hong Kong through grants (16307420, AoE/P-502/20), and the Croucher Foundation (CAS20SC01), the Fund for Shanxi ``1331 Project", Fundamental Research Program of Shanxi Province through 202103021222001. This research was partially conducted using the High Performance Computer of Shanxi University.

\bigskip

\noindent{$^{*}$ These authors contributed equally to this work.\\
\noindent{$^{\dagger)}$zhanglei@sxu.edu.cn}\\
\noindent{$^{\ddagger)}$chenjun@sxu.edu.cn}\\
\noindent{$^{\S)}$phchan@ust.hk}

\bibliography{myref}

\begin{thebibliography}{49}%
\makeatletter
\providecommand \@ifxundefined [1]{%
 \@ifx{#1\undefined}
}%
\providecommand \@ifnum [1]{%
 \ifnum #1\expandafter \@firstoftwo
 \else \expandafter \@secondoftwo
 \fi
}%
\providecommand \@ifx [1]{%
 \ifx #1\expandafter \@firstoftwo
 \else \expandafter \@secondoftwo
 \fi
}%
\providecommand \natexlab [1]{#1}%
\providecommand \enquote  [1]{``#1''}%
\providecommand \bibnamefont  [1]{#1}%
\providecommand \bibfnamefont [1]{#1}%
\providecommand \citenamefont [1]{#1}%
\providecommand \href@noop [0]{\@secondoftwo}%
\providecommand \href [0]{\begingroup \@sanitize@url \@href}%
\providecommand \@href[1]{\@@startlink{#1}\@@href}%
\providecommand \@@href[1]{\endgroup#1\@@endlink}%
\providecommand \@sanitize@url [0]{\catcode `\\12\catcode `\$12\catcode
  `\&12\catcode `\#12\catcode `\^12\catcode `\_12\catcode `\%12\relax}%
\providecommand \@@startlink[1]{}%
\providecommand \@@endlink[0]{}%
\providecommand \url  [0]{\begingroup\@sanitize@url \@url }%
\providecommand \@url [1]{\endgroup\@href {#1}{\urlprefix }}%
\providecommand \urlprefix  [0]{URL }%
\providecommand \Eprint [0]{\href }%
\providecommand \doibase [0]{https://doi.org/}%
\providecommand \selectlanguage [0]{\@gobble}%
\providecommand \bibinfo  [0]{\@secondoftwo}%
\providecommand \bibfield  [0]{\@secondoftwo}%
\providecommand \translation [1]{[#1]}%
\providecommand \BibitemOpen [0]{}%
\providecommand \bibitemStop [0]{}%
\providecommand \bibitemNoStop [0]{.\EOS\space}%
\providecommand \EOS [0]{\spacefactor3000\relax}%
\providecommand \BibitemShut  [1]{\csname bibitem#1\endcsname}%
\let\auto@bib@innerbib\@empty
\bibitem [{\citenamefont {Hasan}\ and\ \citenamefont {Kane}(2010)}]{Hasan1}%
  \BibitemOpen
  \bibfield  {author} {\bibinfo {author} {\bibfnamefont {M.~Z.}\ \bibnamefont
  {Hasan}}\ and\ \bibinfo {author} {\bibfnamefont {C.~L.}\ \bibnamefont
  {Kane}},\ }\bibfield  {title} {\bibinfo {title} {Colloquium: Topological
  insulators},\ }\href {https://doi.org/10.1103/RevModPhys.82.3045} {\bibfield
  {journal} {\bibinfo  {journal} {Rev. Mod. Phys.}\ }\textbf {\bibinfo {volume}
  {82}},\ \bibinfo {pages} {3045} (\bibinfo {year} {2010})}\BibitemShut
  {NoStop}%
\bibitem [{\citenamefont {Qi}\ and\ \citenamefont {Zhang}(2011)}]{Zhang2}%
  \BibitemOpen
  \bibfield  {author} {\bibinfo {author} {\bibfnamefont {X.-L.}\ \bibnamefont
  {Qi}}\ and\ \bibinfo {author} {\bibfnamefont {S.-C.}\ \bibnamefont {Zhang}},\
  }\bibfield  {title} {\bibinfo {title} {Topological insulators and
  superconductors},\ }\href {https://doi.org/10.1103/RevModPhys.83.1057}
  {\bibfield  {journal} {\bibinfo  {journal} {Rev. Mod. Phys.}\ }\textbf
  {\bibinfo {volume} {83}},\ \bibinfo {pages} {1057} (\bibinfo {year}
  {2011})}\BibitemShut {NoStop}%
\bibitem [{\citenamefont {Kane}\ and\ \citenamefont {Mele}(2005)}]{Kane3}%
  \BibitemOpen
  \bibfield  {author} {\bibinfo {author} {\bibfnamefont {C.~L.}\ \bibnamefont
  {Kane}}\ and\ \bibinfo {author} {\bibfnamefont {E.~J.}\ \bibnamefont
  {Mele}},\ }\bibfield  {title} {\bibinfo {title} {Quantum spin {H}all effect
  in graphene},\ }\href {https://doi.org/10.1103/PhysRevLett.95.226801}
  {\bibfield  {journal} {\bibinfo  {journal} {Phys. Rev. Lett.}\ }\textbf
  {\bibinfo {volume} {95}},\ \bibinfo {pages} {226801} (\bibinfo {year}
  {2005})}\BibitemShut {NoStop}%
\bibitem [{\citenamefont {Bernevig}\ and\ \citenamefont
  {Zhang}(2006)}]{Zhang4}%
  \BibitemOpen
  \bibfield  {author} {\bibinfo {author} {\bibfnamefont {B.~A.}\ \bibnamefont
  {Bernevig}}\ and\ \bibinfo {author} {\bibfnamefont {S.-C.}\ \bibnamefont
  {Zhang}},\ }\bibfield  {title} {\bibinfo {title} {Quantum spin {H}all
  effect},\ }\href {https://doi.org/10.1103/PhysRevLett.96.106802} {\bibfield
  {journal} {\bibinfo  {journal} {Phys. Rev. Lett.}\ }\textbf {\bibinfo
  {volume} {96}},\ \bibinfo {pages} {106802} (\bibinfo {year}
  {2006})}\BibitemShut {NoStop}%
\bibitem [{\citenamefont {Ozawa}\ \emph {et~al.}(2019)\citenamefont {Ozawa},
  \citenamefont {Price}, \citenamefont {Amo}, \citenamefont {Goldman},
  \citenamefont {Hafezi}, \citenamefont {Lu}, \citenamefont {Rechtsman},
  \citenamefont {Schuster}, \citenamefont {Simon}, \citenamefont {Zilberberg},\
  and\ \citenamefont {Carusotto}}]{Ozawa5}%
  \BibitemOpen
  \bibfield  {author} {\bibinfo {author} {\bibfnamefont {T.}~\bibnamefont
  {Ozawa}}, \bibinfo {author} {\bibfnamefont {H.~M.}\ \bibnamefont {Price}},
  \bibinfo {author} {\bibfnamefont {A.}~\bibnamefont {Amo}}, \bibinfo {author}
  {\bibfnamefont {N.}~\bibnamefont {Goldman}}, \bibinfo {author} {\bibfnamefont
  {M.}~\bibnamefont {Hafezi}}, \bibinfo {author} {\bibfnamefont
  {L.}~\bibnamefont {Lu}}, \bibinfo {author} {\bibfnamefont {M.~C.}\
  \bibnamefont {Rechtsman}}, \bibinfo {author} {\bibfnamefont {D.}~\bibnamefont
  {Schuster}}, \bibinfo {author} {\bibfnamefont {J.}~\bibnamefont {Simon}},
  \bibinfo {author} {\bibfnamefont {O.}~\bibnamefont {Zilberberg}},\ and\
  \bibinfo {author} {\bibfnamefont {I.}~\bibnamefont {Carusotto}},\ }\bibfield
  {title} {\bibinfo {title} {Topological photonics},\ }\href
  {https://doi.org/10.1103/RevModPhys.91.015006} {\bibfield  {journal}
  {\bibinfo  {journal} {Rev. Mod. Phys.}\ }\textbf {\bibinfo {volume} {91}},\
  \bibinfo {pages} {015006} (\bibinfo {year} {2019})}\BibitemShut {NoStop}%
\bibitem [{\citenamefont {Lu}\ \emph {et~al.}(2014)\citenamefont {Lu},
  \citenamefont {Joannopoulos},\ and\ \citenamefont
  {Solja{\v{c}}i{\'{c}}}}]{Lu6}%
  \BibitemOpen
  \bibfield  {author} {\bibinfo {author} {\bibfnamefont {L.}~\bibnamefont
  {Lu}}, \bibinfo {author} {\bibfnamefont {J.~D.}\ \bibnamefont
  {Joannopoulos}},\ and\ \bibinfo {author} {\bibfnamefont {M.}~\bibnamefont
  {Solja{\v{c}}i{\'{c}}}},\ }\bibfield  {title} {\bibinfo {title} {Topological
  photonics},\ }\href {https://doi.org/10.1038/nphoton.2014.248} {\bibfield
  {journal} {\bibinfo  {journal} {Nat. Photonics}\ }\textbf {\bibinfo {volume}
  {8}},\ \bibinfo {pages} {821} (\bibinfo {year} {2014})}\BibitemShut {NoStop}%
\bibitem [{\citenamefont {Kim}\ \emph {et~al.}(2020)\citenamefont {Kim},
  \citenamefont {Jacob},\ and\ \citenamefont {Rho}}]{Kim7}%
  \BibitemOpen
  \bibfield  {author} {\bibinfo {author} {\bibfnamefont {M.}~\bibnamefont
  {Kim}}, \bibinfo {author} {\bibfnamefont {Z.}~\bibnamefont {Jacob}},\ and\
  \bibinfo {author} {\bibfnamefont {J.}~\bibnamefont {Rho}},\ }\bibfield
  {title} {\bibinfo {title} {Recent advances in 2d, 3d and higher-order
  topological photonics},\ }\href {https://doi.org/10.1038/s41377-020-0331-y}
  {\bibfield  {journal} {\bibinfo  {journal} {Light Sci. Appl.}\ }\textbf
  {\bibinfo {volume} {9}},\ \bibinfo {pages} {130} (\bibinfo {year}
  {2020})}\BibitemShut {NoStop}%
\bibitem [{\citenamefont {Haldane}\ and\ \citenamefont
  {Raghu}(2008)}]{Haldane8}%
  \BibitemOpen
  \bibfield  {author} {\bibinfo {author} {\bibfnamefont {F.~D.~M.}\
  \bibnamefont {Haldane}}\ and\ \bibinfo {author} {\bibfnamefont
  {S.}~\bibnamefont {Raghu}},\ }\bibfield  {title} {\bibinfo {title} {Possible
  realization of directional optical waveguides in photonic crystals with
  broken time-reversal symmetry},\ }\href
  {https://doi.org/10.1103/PhysRevLett.100.013904} {\bibfield  {journal}
  {\bibinfo  {journal} {Phys. Rev. Lett.}\ }\textbf {\bibinfo {volume} {100}},\
  \bibinfo {pages} {013904} (\bibinfo {year} {2008})}\BibitemShut {NoStop}%
\bibitem [{\citenamefont {Wang}\ \emph {et~al.}(2009)\citenamefont {Wang},
  \citenamefont {Chong}, \citenamefont {Joannopoulos},\ and\ \citenamefont
  {Soljačić}}]{Wang9}%
  \BibitemOpen
  \bibfield  {author} {\bibinfo {author} {\bibfnamefont {Z.}~\bibnamefont
  {Wang}}, \bibinfo {author} {\bibfnamefont {Y.}~\bibnamefont {Chong}},
  \bibinfo {author} {\bibfnamefont {J.~D.}\ \bibnamefont {Joannopoulos}},\ and\
  \bibinfo {author} {\bibfnamefont {M.}~\bibnamefont {Soljačić}},\ }\bibfield
   {title} {\bibinfo {title} {Observation of unidirectional
  backscattering-immune topological electromagnetic states},\ }\href
  {https://doi.org/10.1038/nature08293} {\bibfield  {journal} {\bibinfo
  {journal} {Nature}\ }\textbf {\bibinfo {volume} {461}},\ \bibinfo {pages}
  {772} (\bibinfo {year} {2009})}\BibitemShut {NoStop}%
\bibitem [{\citenamefont {Wu}\ and\ \citenamefont {Hu}(2015)}]{Hu10}%
  \BibitemOpen
  \bibfield  {author} {\bibinfo {author} {\bibfnamefont {L.-H.}\ \bibnamefont
  {Wu}}\ and\ \bibinfo {author} {\bibfnamefont {X.}~\bibnamefont {Hu}},\
  }\bibfield  {title} {\bibinfo {title} {Scheme for achieving a topological
  photonic crystal by using dielectric material},\ }\href
  {https://doi.org/10.1103/PhysRevLett.114.223901} {\bibfield  {journal}
  {\bibinfo  {journal} {Phys. Rev. Lett.}\ }\textbf {\bibinfo {volume} {114}},\
  \bibinfo {pages} {223901} (\bibinfo {year} {2015})}\BibitemShut {NoStop}%
\bibitem [{\citenamefont {Barik}\ \emph {et~al.}(2018)\citenamefont {Barik},
  \citenamefont {Karasahin}, \citenamefont {Flower}, \citenamefont {Cai},
  \citenamefont {Miyake}, \citenamefont {DeGottardi}, \citenamefont {Hafezi},\
  and\ \citenamefont {Waks}}]{Barik11}%
  \BibitemOpen
  \bibfield  {author} {\bibinfo {author} {\bibfnamefont {S.}~\bibnamefont
  {Barik}}, \bibinfo {author} {\bibfnamefont {A.}~\bibnamefont {Karasahin}},
  \bibinfo {author} {\bibfnamefont {C.}~\bibnamefont {Flower}}, \bibinfo
  {author} {\bibfnamefont {T.}~\bibnamefont {Cai}}, \bibinfo {author}
  {\bibfnamefont {H.}~\bibnamefont {Miyake}}, \bibinfo {author} {\bibfnamefont
  {W.}~\bibnamefont {DeGottardi}}, \bibinfo {author} {\bibfnamefont
  {M.}~\bibnamefont {Hafezi}},\ and\ \bibinfo {author} {\bibfnamefont
  {E.}~\bibnamefont {Waks}},\ }\bibfield  {title} {\bibinfo {title} {A
  topological quantum optics interface},\ }\href
  {https://doi.org/10.1126/science.aaq0327} {\bibfield  {journal} {\bibinfo
  {journal} {Science}\ }\textbf {\bibinfo {volume} {359}},\ \bibinfo {pages}
  {666} (\bibinfo {year} {2018})}\BibitemShut {NoStop}%
\bibitem [{\citenamefont {Shalaev}\ \emph {et~al.}(2019)\citenamefont
  {Shalaev}, \citenamefont {Walasik}, \citenamefont {Tsukernik}, \citenamefont
  {Xu},\ and\ \citenamefont {Litchinitser}}]{Shalaev12}%
  \BibitemOpen
  \bibfield  {author} {\bibinfo {author} {\bibfnamefont {M.~I.}\ \bibnamefont
  {Shalaev}}, \bibinfo {author} {\bibfnamefont {W.}~\bibnamefont {Walasik}},
  \bibinfo {author} {\bibfnamefont {A.}~\bibnamefont {Tsukernik}}, \bibinfo
  {author} {\bibfnamefont {Y.}~\bibnamefont {Xu}},\ and\ \bibinfo {author}
  {\bibfnamefont {N.~M.}\ \bibnamefont {Litchinitser}},\ }\bibfield  {title}
  {\bibinfo {title} {Robust topologically protected transport in photonic
  crystals at telecommunication wavelengths},\ }\href
  {https://doi.org/10.1038/s41565-018-0297-6} {\bibfield  {journal} {\bibinfo
  {journal} {Nat. Nanotechnol.}\ }\textbf {\bibinfo {volume} {14}},\ \bibinfo
  {pages} {31} (\bibinfo {year} {2019})}\BibitemShut {NoStop}%
\bibitem [{\citenamefont {Yang}\ \emph {et~al.}(2020)\citenamefont {Yang},
  \citenamefont {Yamagami}, \citenamefont {Yu}, \citenamefont {Pitchappa},
  \citenamefont {Webber}, \citenamefont {Zhang}, \citenamefont {Fujita},
  \citenamefont {Nagatsuma},\ and\ \citenamefont {Singh}}]{Yang13}%
  \BibitemOpen
  \bibfield  {author} {\bibinfo {author} {\bibfnamefont {Y.}~\bibnamefont
  {Yang}}, \bibinfo {author} {\bibfnamefont {Y.}~\bibnamefont {Yamagami}},
  \bibinfo {author} {\bibfnamefont {X.}~\bibnamefont {Yu}}, \bibinfo {author}
  {\bibfnamefont {P.}~\bibnamefont {Pitchappa}}, \bibinfo {author}
  {\bibfnamefont {J.}~\bibnamefont {Webber}}, \bibinfo {author} {\bibfnamefont
  {B.}~\bibnamefont {Zhang}}, \bibinfo {author} {\bibfnamefont
  {M.}~\bibnamefont {Fujita}}, \bibinfo {author} {\bibfnamefont
  {T.}~\bibnamefont {Nagatsuma}},\ and\ \bibinfo {author} {\bibfnamefont
  {R.}~\bibnamefont {Singh}},\ }\bibfield  {title} {\bibinfo {title} {Terahertz
  topological photonics for on-chip communication},\ }\href
  {https://doi.org/10.1038/s41566-020-0618-9} {\bibfield  {journal} {\bibinfo
  {journal} {Nat. Photonics}\ }\textbf {\bibinfo {volume} {14}},\ \bibinfo
  {pages} {446} (\bibinfo {year} {2020})}\BibitemShut {NoStop}%
\bibitem [{\citenamefont {Wang}\ \emph {et~al.}(2021)\citenamefont {Wang},
  \citenamefont {Zhang}, \citenamefont {Zhang}, \citenamefont {Wang},
  \citenamefont {Guo}, \citenamefont {Zhang},\ and\ \citenamefont
  {Chan}}]{Wang14}%
  \BibitemOpen
  \bibfield  {author} {\bibinfo {author} {\bibfnamefont {M.}~\bibnamefont
  {Wang}}, \bibinfo {author} {\bibfnamefont {R.-Y.}\ \bibnamefont {Zhang}},
  \bibinfo {author} {\bibfnamefont {L.}~\bibnamefont {Zhang}}, \bibinfo
  {author} {\bibfnamefont {D.}~\bibnamefont {Wang}}, \bibinfo {author}
  {\bibfnamefont {Q.}~\bibnamefont {Guo}}, \bibinfo {author} {\bibfnamefont
  {Z.-Q.}\ \bibnamefont {Zhang}},\ and\ \bibinfo {author} {\bibfnamefont
  {C.~T.}\ \bibnamefont {Chan}},\ }\bibfield  {title} {\bibinfo {title}
  {Topological one-way large-area waveguide states in magnetic photonic
  crystals},\ }\href {https://doi.org/10.1103/PhysRevLett.126.067401}
  {\bibfield  {journal} {\bibinfo  {journal} {Phys. Rev. Lett.}\ }\textbf
  {\bibinfo {volume} {126}},\ \bibinfo {pages} {067401} (\bibinfo {year}
  {2021})}\BibitemShut {NoStop}%
\bibitem [{\citenamefont {Wang}\ \emph {et~al.}(2015)\citenamefont {Wang},
  \citenamefont {Lu},\ and\ \citenamefont {Bertoldi}}]{Pai15}%
  \BibitemOpen
  \bibfield  {author} {\bibinfo {author} {\bibfnamefont {P.}~\bibnamefont
  {Wang}}, \bibinfo {author} {\bibfnamefont {L.}~\bibnamefont {Lu}},\ and\
  \bibinfo {author} {\bibfnamefont {K.}~\bibnamefont {Bertoldi}},\ }\bibfield
  {title} {\bibinfo {title} {Topological phononic crystals with one-way elastic
  edge waves},\ }\href {https://doi.org/10.1103/PhysRevLett.115.104302}
  {\bibfield  {journal} {\bibinfo  {journal} {Phys. Rev. Lett.}\ }\textbf
  {\bibinfo {volume} {115}},\ \bibinfo {pages} {104302} (\bibinfo {year}
  {2015})}\BibitemShut {NoStop}%
\bibitem [{\citenamefont {Khanikaev}\ \emph {et~al.}(2015)\citenamefont
  {Khanikaev}, \citenamefont {Fleury}, \citenamefont {Mousavi},\ and\
  \citenamefont {Alù}}]{Khanikaev16}%
  \BibitemOpen
  \bibfield  {author} {\bibinfo {author} {\bibfnamefont {A.~B.}\ \bibnamefont
  {Khanikaev}}, \bibinfo {author} {\bibfnamefont {R.}~\bibnamefont {Fleury}},
  \bibinfo {author} {\bibfnamefont {S.~H.}\ \bibnamefont {Mousavi}},\ and\
  \bibinfo {author} {\bibfnamefont {A.}~\bibnamefont {Alù}},\ }\bibfield
  {title} {\bibinfo {title} {Topologically robust sound propagation in an
  angular-momentum-biased graphene-like resonator lattice},\ }\href
  {https://doi.org/10.1038/ncomms9260} {\bibfield  {journal} {\bibinfo
  {journal} {Nat. Commun.}\ }\textbf {\bibinfo {volume} {6}},\ \bibinfo {pages}
  {8260} (\bibinfo {year} {2015})}\BibitemShut {NoStop}%
\bibitem [{\citenamefont {Yang}\ \emph {et~al.}(2015)\citenamefont {Yang},
  \citenamefont {Gao}, \citenamefont {Shi}, \citenamefont {Lin}, \citenamefont
  {Gao}, \citenamefont {Chong},\ and\ \citenamefont {Zhang}}]{Zhaoju17}%
  \BibitemOpen
  \bibfield  {author} {\bibinfo {author} {\bibfnamefont {Z.}~\bibnamefont
  {Yang}}, \bibinfo {author} {\bibfnamefont {F.}~\bibnamefont {Gao}}, \bibinfo
  {author} {\bibfnamefont {X.}~\bibnamefont {Shi}}, \bibinfo {author}
  {\bibfnamefont {X.}~\bibnamefont {Lin}}, \bibinfo {author} {\bibfnamefont
  {Z.}~\bibnamefont {Gao}}, \bibinfo {author} {\bibfnamefont {Y.}~\bibnamefont
  {Chong}},\ and\ \bibinfo {author} {\bibfnamefont {B.}~\bibnamefont {Zhang}},\
  }\bibfield  {title} {\bibinfo {title} {Topological acoustics},\ }\href
  {https://doi.org/10.1103/PhysRevLett.114.114301} {\bibfield  {journal}
  {\bibinfo  {journal} {Phys. Rev. Lett.}\ }\textbf {\bibinfo {volume} {114}},\
  \bibinfo {pages} {114301} (\bibinfo {year} {2015})}\BibitemShut {NoStop}%
\bibitem [{\citenamefont {He}\ \emph {et~al.}(2016)\citenamefont {He},
  \citenamefont {Ni}, \citenamefont {Ge}, \citenamefont {Sun}, \citenamefont
  {Chen}, \citenamefont {Lu}, \citenamefont {Liu},\ and\ \citenamefont
  {Chen}}]{He18}%
  \BibitemOpen
  \bibfield  {author} {\bibinfo {author} {\bibfnamefont {C.}~\bibnamefont
  {He}}, \bibinfo {author} {\bibfnamefont {X.}~\bibnamefont {Ni}}, \bibinfo
  {author} {\bibfnamefont {H.}~\bibnamefont {Ge}}, \bibinfo {author}
  {\bibfnamefont {X.-C.}\ \bibnamefont {Sun}}, \bibinfo {author} {\bibfnamefont
  {Y.-B.}\ \bibnamefont {Chen}}, \bibinfo {author} {\bibfnamefont {M.-H.}\
  \bibnamefont {Lu}}, \bibinfo {author} {\bibfnamefont {X.-P.}\ \bibnamefont
  {Liu}},\ and\ \bibinfo {author} {\bibfnamefont {Y.-F.}\ \bibnamefont
  {Chen}},\ }\bibfield  {title} {\bibinfo {title} {Acoustic topological
  insulator and robust one-way sound transport},\ }\href
  {https://doi.org/10.1038/nphys3867} {\bibfield  {journal} {\bibinfo
  {journal} {Nat. Phys.}\ }\textbf {\bibinfo {volume} {12}},\ \bibinfo {pages}
  {1124} (\bibinfo {year} {2016})}\BibitemShut {NoStop}%
\bibitem [{\citenamefont {Lu}\ \emph {et~al.}(2017)\citenamefont {Lu},
  \citenamefont {Qiu}, \citenamefont {Ye}, \citenamefont {Fan}, \citenamefont
  {Ke}, \citenamefont {Zhang},\ and\ \citenamefont {Liu}}]{Jiuyang19}%
  \BibitemOpen
  \bibfield  {author} {\bibinfo {author} {\bibfnamefont {J.}~\bibnamefont
  {Lu}}, \bibinfo {author} {\bibfnamefont {C.}~\bibnamefont {Qiu}}, \bibinfo
  {author} {\bibfnamefont {L.}~\bibnamefont {Ye}}, \bibinfo {author}
  {\bibfnamefont {X.}~\bibnamefont {Fan}}, \bibinfo {author} {\bibfnamefont
  {M.}~\bibnamefont {Ke}}, \bibinfo {author} {\bibfnamefont {F.}~\bibnamefont
  {Zhang}},\ and\ \bibinfo {author} {\bibfnamefont {Z.}~\bibnamefont {Liu}},\
  }\bibfield  {title} {\bibinfo {title} {Observation of topological valley
  transport of sound in sonic crystals},\ }\href
  {https://doi.org/10.1038/nphys3999} {\bibfield  {journal} {\bibinfo
  {journal} {Nat. Phys.}\ }\textbf {\bibinfo {volume} {13}},\ \bibinfo {pages}
  {369} (\bibinfo {year} {2017})}\BibitemShut {NoStop}%
\bibitem [{\citenamefont {Ding}\ \emph {et~al.}(2019)\citenamefont {Ding},
  \citenamefont {Peng}, \citenamefont {Zhu}, \citenamefont {Fan}, \citenamefont
  {Yang}, \citenamefont {Liang}, \citenamefont {Zhu}, \citenamefont {Wan},\
  and\ \citenamefont {Cheng}}]{Ding20}%
  \BibitemOpen
  \bibfield  {author} {\bibinfo {author} {\bibfnamefont {Y.}~\bibnamefont
  {Ding}}, \bibinfo {author} {\bibfnamefont {Y.}~\bibnamefont {Peng}}, \bibinfo
  {author} {\bibfnamefont {Y.}~\bibnamefont {Zhu}}, \bibinfo {author}
  {\bibfnamefont {X.}~\bibnamefont {Fan}}, \bibinfo {author} {\bibfnamefont
  {J.}~\bibnamefont {Yang}}, \bibinfo {author} {\bibfnamefont {B.}~\bibnamefont
  {Liang}}, \bibinfo {author} {\bibfnamefont {X.}~\bibnamefont {Zhu}}, \bibinfo
  {author} {\bibfnamefont {X.}~\bibnamefont {Wan}},\ and\ \bibinfo {author}
  {\bibfnamefont {J.}~\bibnamefont {Cheng}},\ }\bibfield  {title} {\bibinfo
  {title} {Experimental demonstration of acoustic chern insulators},\ }\href
  {https://doi.org/10.1103/PhysRevLett.122.014302} {\bibfield  {journal}
  {\bibinfo  {journal} {Phys. Rev. Lett.}\ }\textbf {\bibinfo {volume} {122}},\
  \bibinfo {pages} {014302} (\bibinfo {year} {2019})}\BibitemShut {NoStop}%
\bibitem [{\citenamefont {Wang}\ \emph {et~al.}(2020)\citenamefont {Wang},
  \citenamefont {Zhou}, \citenamefont {Bi}, \citenamefont {Qiu}, \citenamefont
  {Ke},\ and\ \citenamefont {Liu}}]{Mudi21}%
  \BibitemOpen
  \bibfield  {author} {\bibinfo {author} {\bibfnamefont {M.}~\bibnamefont
  {Wang}}, \bibinfo {author} {\bibfnamefont {W.}~\bibnamefont {Zhou}}, \bibinfo
  {author} {\bibfnamefont {L.}~\bibnamefont {Bi}}, \bibinfo {author}
  {\bibfnamefont {C.}~\bibnamefont {Qiu}}, \bibinfo {author} {\bibfnamefont
  {M.}~\bibnamefont {Ke}},\ and\ \bibinfo {author} {\bibfnamefont
  {Z.}~\bibnamefont {Liu}},\ }\bibfield  {title} {\bibinfo {title}
  {Valley-locked waveguide transport in acoustic heterostructures},\ }\href
  {https://doi.org/10.1038/s41467-020-16843-z} {\bibfield  {journal} {\bibinfo
  {journal} {Nat. Commun.}\ }\textbf {\bibinfo {volume} {11}},\ \bibinfo
  {pages} {3000} (\bibinfo {year} {2020})}\BibitemShut {NoStop}%
\bibitem [{\citenamefont {Ma}\ \emph {et~al.}(2019)\citenamefont {Ma},
  \citenamefont {Xiao},\ and\ \citenamefont {Chan}}]{Guancong22}%
  \BibitemOpen
  \bibfield  {author} {\bibinfo {author} {\bibfnamefont {G.}~\bibnamefont
  {Ma}}, \bibinfo {author} {\bibfnamefont {M.}~\bibnamefont {Xiao}},\ and\
  \bibinfo {author} {\bibfnamefont {C.~T.}\ \bibnamefont {Chan}},\ }\bibfield
  {title} {\bibinfo {title} {Topological phases in acoustic and mechanical
  systems},\ }\href {https://doi.org/10.1038/s42254-019-0030-x} {\bibfield
  {journal} {\bibinfo  {journal} {Nat. Rev. Phys.}\ }\textbf {\bibinfo {volume}
  {1}},\ \bibinfo {pages} {281} (\bibinfo {year} {2019})}\BibitemShut {NoStop}%
\bibitem [{\citenamefont {Xue}\ \emph {et~al.}(2022)\citenamefont {Xue},
  \citenamefont {Yang},\ and\ \citenamefont {Zhang}}]{Haoran23}%
  \BibitemOpen
  \bibfield  {author} {\bibinfo {author} {\bibfnamefont {H.}~\bibnamefont
  {Xue}}, \bibinfo {author} {\bibfnamefont {Y.}~\bibnamefont {Yang}},\ and\
  \bibinfo {author} {\bibfnamefont {B.}~\bibnamefont {Zhang}},\ }\bibfield
  {title} {\bibinfo {title} {Topological acoustics},\ }\href
  {https://doi.org/10.1038/s41578-022-00465-6} {\bibfield  {journal} {\bibinfo
  {journal} {Nat. Rev. Mater.}\ }\textbf {\bibinfo {volume} {7}},\ \bibinfo
  {pages} {974} (\bibinfo {year} {2022})}\BibitemShut {NoStop}%
\bibitem [{\citenamefont {Thouless}\ \emph {et~al.}(1982)\citenamefont
  {Thouless}, \citenamefont {Kohmoto}, \citenamefont {Nightingale},\ and\
  \citenamefont {Den~Nijs}}]{Thouless24}%
  \BibitemOpen
  \bibfield  {author} {\bibinfo {author} {\bibfnamefont {D.~J.}\ \bibnamefont
  {Thouless}}, \bibinfo {author} {\bibfnamefont {M.}~\bibnamefont {Kohmoto}},
  \bibinfo {author} {\bibfnamefont {M.~P.}\ \bibnamefont {Nightingale}},\ and\
  \bibinfo {author} {\bibfnamefont {M.}~\bibnamefont {Den~Nijs}},\ }\bibfield
  {title} {\bibinfo {title} {Quantized {H}all conductance in a two-dimensional
  periodic potential},\ }\href {https://doi.org/10.1103/PhysRevLett.49.405}
  {\bibfield  {journal} {\bibinfo  {journal} {Phys. Rev. Lett.}\ }\textbf
  {\bibinfo {volume} {49}},\ \bibinfo {pages} {405} (\bibinfo {year}
  {1982})}\BibitemShut {NoStop}%
\bibitem [{\citenamefont {Mitchell}\ \emph {et~al.}(2018)\citenamefont
  {Mitchell}, \citenamefont {Nash}, \citenamefont {Hexner}, \citenamefont
  {Turner},\ and\ \citenamefont {Irvine}}]{Mitchell25}%
  \BibitemOpen
  \bibfield  {author} {\bibinfo {author} {\bibfnamefont {N.~P.}\ \bibnamefont
  {Mitchell}}, \bibinfo {author} {\bibfnamefont {L.~M.}\ \bibnamefont {Nash}},
  \bibinfo {author} {\bibfnamefont {D.}~\bibnamefont {Hexner}}, \bibinfo
  {author} {\bibfnamefont {A.~M.}\ \bibnamefont {Turner}},\ and\ \bibinfo
  {author} {\bibfnamefont {W.~T.~M.}\ \bibnamefont {Irvine}},\ }\bibfield
  {title} {\bibinfo {title} {Amorphous topological insulators constructed from
  random point sets},\ }\href {https://doi.org/10.1038/s41567-017-0024-5}
  {\bibfield  {journal} {\bibinfo  {journal} {Nat. Phys.}\ }\textbf {\bibinfo
  {volume} {14}},\ \bibinfo {pages} {380} (\bibinfo {year} {2018})}\BibitemShut
  {NoStop}%
\bibitem [{\citenamefont {Zhang}\ \emph {et~al.}(2023)\citenamefont {Zhang},
  \citenamefont {Delplace},\ and\ \citenamefont {Fleury}}]{Zhe26}%
  \BibitemOpen
  \bibfield  {author} {\bibinfo {author} {\bibfnamefont {Z.}~\bibnamefont
  {Zhang}}, \bibinfo {author} {\bibfnamefont {P.}~\bibnamefont {Delplace}},\
  and\ \bibinfo {author} {\bibfnamefont {R.}~\bibnamefont {Fleury}},\
  }\bibfield  {title} {\bibinfo {title} {Anomalous topological waves in
  strongly amorphous scattering networks},\ }\href
  {https://doi.org/10.1126/sciadv.adg3186} {\bibfield  {journal} {\bibinfo
  {journal} {Sci. Adv.}\ }\textbf {\bibinfo {volume} {9}},\ \bibinfo {pages}
  {eadg3186} (\bibinfo {year} {2023})}\BibitemShut {NoStop}%
\bibitem [{\citenamefont {Agarwala}\ and\ \citenamefont
  {Shenoy}(2017)}]{Agarwala27}%
  \BibitemOpen
  \bibfield  {author} {\bibinfo {author} {\bibfnamefont {A.}~\bibnamefont
  {Agarwala}}\ and\ \bibinfo {author} {\bibfnamefont {V.~B.}\ \bibnamefont
  {Shenoy}},\ }\bibfield  {title} {\bibinfo {title} {Topological insulators in
  amorphous systems},\ }\href {https://doi.org/10.1103/PhysRevLett.118.236402}
  {\bibfield  {journal} {\bibinfo  {journal} {Phys. Rev. Lett.}\ }\textbf
  {\bibinfo {volume} {118}},\ \bibinfo {pages} {236402} (\bibinfo {year}
  {2017})}\BibitemShut {NoStop}%
\bibitem [{\citenamefont {Yang}\ \emph
  {et~al.}(2019{\natexlab{a}})\citenamefont {Yang}, \citenamefont {Qin},
  \citenamefont {Deng}, \citenamefont {Duan},\ and\ \citenamefont
  {Xu}}]{Yang28}%
  \BibitemOpen
  \bibfield  {author} {\bibinfo {author} {\bibfnamefont {Y.-B.}\ \bibnamefont
  {Yang}}, \bibinfo {author} {\bibfnamefont {T.}~\bibnamefont {Qin}}, \bibinfo
  {author} {\bibfnamefont {D.-L.}\ \bibnamefont {Deng}}, \bibinfo {author}
  {\bibfnamefont {L.-M.}\ \bibnamefont {Duan}},\ and\ \bibinfo {author}
  {\bibfnamefont {Y.}~\bibnamefont {Xu}},\ }\bibfield  {title} {\bibinfo
  {title} {Topological amorphous metals},\ }\href
  {https://doi.org/10.1103/PhysRevLett.123.076401} {\bibfield  {journal}
  {\bibinfo  {journal} {Phys. Rev. Lett.}\ }\textbf {\bibinfo {volume} {123}},\
  \bibinfo {pages} {076401} (\bibinfo {year} {2019}{\natexlab{a}})}\BibitemShut
  {NoStop}%
\bibitem [{\citenamefont {Wang}\ \emph {et~al.}(2022)\citenamefont {Wang},
  \citenamefont {Cheng}, \citenamefont {Liu}, \citenamefont {Liu},\ and\
  \citenamefont {Huang}}]{Wang29}%
  \BibitemOpen
  \bibfield  {author} {\bibinfo {author} {\bibfnamefont {C.}~\bibnamefont
  {Wang}}, \bibinfo {author} {\bibfnamefont {T.}~\bibnamefont {Cheng}},
  \bibinfo {author} {\bibfnamefont {Z.}~\bibnamefont {Liu}}, \bibinfo {author}
  {\bibfnamefont {F.}~\bibnamefont {Liu}},\ and\ \bibinfo {author}
  {\bibfnamefont {H.}~\bibnamefont {Huang}},\ }\bibfield  {title} {\bibinfo
  {title} {Structural amorphization-induced topological order},\ }\href
  {https://doi.org/10.1103/PhysRevLett.128.056401} {\bibfield  {journal}
  {\bibinfo  {journal} {Phys. Rev. Lett.}\ }\textbf {\bibinfo {volume} {128}},\
  \bibinfo {pages} {056401} (\bibinfo {year} {2022})}\BibitemShut {NoStop}%
\bibitem [{\citenamefont {Ivaki}\ \emph {et~al.}(2020)\citenamefont {Ivaki},
  \citenamefont {Sahlberg},\ and\ \citenamefont {Ojanen}}]{Ivaki30}%
  \BibitemOpen
  \bibfield  {author} {\bibinfo {author} {\bibfnamefont {M.~N.}\ \bibnamefont
  {Ivaki}}, \bibinfo {author} {\bibfnamefont {I.}~\bibnamefont {Sahlberg}},\
  and\ \bibinfo {author} {\bibfnamefont {T.}~\bibnamefont {Ojanen}},\
  }\bibfield  {title} {\bibinfo {title} {Criticality in amorphous topological
  matter: Beyond the universal scaling paradigm},\ }\href
  {https://doi.org/10.1103/PhysRevResearch.2.043301} {\bibfield  {journal}
  {\bibinfo  {journal} {Phys. Rev. Res.}\ }\textbf {\bibinfo {volume} {2}},\
  \bibinfo {pages} {043301} (\bibinfo {year} {2020})}\BibitemShut {NoStop}%
\bibitem [{\citenamefont {Agarwala}\ \emph {et~al.}(2020)\citenamefont
  {Agarwala}, \citenamefont {Juri\ifmmode \check{c}\else
  \v{c}\fi{}i\ifmmode~\acute{c}\else \'{c}\fi{}},\ and\ \citenamefont
  {Roy}}]{Agarwala31}%
  \BibitemOpen
  \bibfield  {author} {\bibinfo {author} {\bibfnamefont {A.}~\bibnamefont
  {Agarwala}}, \bibinfo {author} {\bibfnamefont {V.}~\bibnamefont {Juri\ifmmode
  \check{c}\else \v{c}\fi{}i\ifmmode~\acute{c}\else \'{c}\fi{}}},\ and\
  \bibinfo {author} {\bibfnamefont {B.}~\bibnamefont {Roy}},\ }\bibfield
  {title} {\bibinfo {title} {Higher-order topological insulators in amorphous
  solids},\ }\href {https://doi.org/10.1103/PhysRevResearch.2.012067}
  {\bibfield  {journal} {\bibinfo  {journal} {Phys. Rev. Res.}\ }\textbf
  {\bibinfo {volume} {2}},\ \bibinfo {pages} {012067} (\bibinfo {year}
  {2020})}\BibitemShut {NoStop}%
\bibitem [{\citenamefont {Kraus}\ \emph {et~al.}(2012)\citenamefont {Kraus},
  \citenamefont {Lahini}, \citenamefont {Ringel}, \citenamefont {Verbin},\ and\
  \citenamefont {Zilberberg}}]{Kraus32}%
  \BibitemOpen
  \bibfield  {author} {\bibinfo {author} {\bibfnamefont {Y.~E.}\ \bibnamefont
  {Kraus}}, \bibinfo {author} {\bibfnamefont {Y.}~\bibnamefont {Lahini}},
  \bibinfo {author} {\bibfnamefont {Z.}~\bibnamefont {Ringel}}, \bibinfo
  {author} {\bibfnamefont {M.}~\bibnamefont {Verbin}},\ and\ \bibinfo {author}
  {\bibfnamefont {O.}~\bibnamefont {Zilberberg}},\ }\bibfield  {title}
  {\bibinfo {title} {Topological states and adiabatic pumping in
  quasicrystals},\ }\href {https://doi.org/10.1103/PhysRevLett.109.106402}
  {\bibfield  {journal} {\bibinfo  {journal} {Phys. Rev. Lett.}\ }\textbf
  {\bibinfo {volume} {109}},\ \bibinfo {pages} {106402} (\bibinfo {year}
  {2012})}\BibitemShut {NoStop}%
\bibitem [{\citenamefont {Bandres}\ \emph {et~al.}(2016)\citenamefont
  {Bandres}, \citenamefont {Rechtsman},\ and\ \citenamefont
  {Segev}}]{Bandres33}%
  \BibitemOpen
  \bibfield  {author} {\bibinfo {author} {\bibfnamefont {M.~A.}\ \bibnamefont
  {Bandres}}, \bibinfo {author} {\bibfnamefont {M.~C.}\ \bibnamefont
  {Rechtsman}},\ and\ \bibinfo {author} {\bibfnamefont {M.}~\bibnamefont
  {Segev}},\ }\bibfield  {title} {\bibinfo {title} {Topological photonic
  quasicrystals: Fractal topological spectrum and protected transport},\ }\href
  {https://doi.org/10.1103/PhysRevX.6.011016} {\bibfield  {journal} {\bibinfo
  {journal} {Phys. Rev. X}\ }\textbf {\bibinfo {volume} {6}},\ \bibinfo {pages}
  {011016} (\bibinfo {year} {2016})}\BibitemShut {NoStop}%
\bibitem [{\citenamefont {St{\"u}tzer}\ \emph {et~al.}(2018)\citenamefont
  {St{\"u}tzer}, \citenamefont {Plotnik}, \citenamefont {Lumer}, \citenamefont
  {Titum}, \citenamefont {Lindner}, \citenamefont {Segev}, \citenamefont
  {Rechtsman},\ and\ \citenamefont {Szameit}}]{Stutzer34}%
  \BibitemOpen
  \bibfield  {author} {\bibinfo {author} {\bibfnamefont {S.}~\bibnamefont
  {St{\"u}tzer}}, \bibinfo {author} {\bibfnamefont {Y.}~\bibnamefont
  {Plotnik}}, \bibinfo {author} {\bibfnamefont {Y.}~\bibnamefont {Lumer}},
  \bibinfo {author} {\bibfnamefont {P.}~\bibnamefont {Titum}}, \bibinfo
  {author} {\bibfnamefont {N.~H.}\ \bibnamefont {Lindner}}, \bibinfo {author}
  {\bibfnamefont {M.}~\bibnamefont {Segev}}, \bibinfo {author} {\bibfnamefont
  {M.~C.}\ \bibnamefont {Rechtsman}},\ and\ \bibinfo {author} {\bibfnamefont
  {A.}~\bibnamefont {Szameit}},\ }\bibfield  {title} {\bibinfo {title}
  {Photonic topological {A}nderson insulators},\ }\href
  {https://doi.org/10.1038/s41586-018-0418-2} {\bibfield  {journal} {\bibinfo
  {journal} {Nature}\ }\textbf {\bibinfo {volume} {560}},\ \bibinfo {pages}
  {461} (\bibinfo {year} {2018})}\BibitemShut {NoStop}%
\bibitem [{\citenamefont {Liu}\ \emph {et~al.}(2020)\citenamefont {Liu},
  \citenamefont {Yang}, \citenamefont {Ren}, \citenamefont {Xue}, \citenamefont
  {Lin}, \citenamefont {Hu}, \citenamefont {Sun}, \citenamefont {Peng},
  \citenamefont {Zhou}, \citenamefont {Chong},\ and\ \citenamefont
  {Zhang}}]{Liu35}%
  \BibitemOpen
  \bibfield  {author} {\bibinfo {author} {\bibfnamefont {G.-G.}\ \bibnamefont
  {Liu}}, \bibinfo {author} {\bibfnamefont {Y.}~\bibnamefont {Yang}}, \bibinfo
  {author} {\bibfnamefont {X.}~\bibnamefont {Ren}}, \bibinfo {author}
  {\bibfnamefont {H.}~\bibnamefont {Xue}}, \bibinfo {author} {\bibfnamefont
  {X.}~\bibnamefont {Lin}}, \bibinfo {author} {\bibfnamefont {Y.-H.}\
  \bibnamefont {Hu}}, \bibinfo {author} {\bibfnamefont {H.-X.}\ \bibnamefont
  {Sun}}, \bibinfo {author} {\bibfnamefont {B.}~\bibnamefont {Peng}}, \bibinfo
  {author} {\bibfnamefont {P.}~\bibnamefont {Zhou}}, \bibinfo {author}
  {\bibfnamefont {Y.}~\bibnamefont {Chong}},\ and\ \bibinfo {author}
  {\bibfnamefont {B.}~\bibnamefont {Zhang}},\ }\bibfield  {title} {\bibinfo
  {title} {Topological {A}nderson insulator in disordered photonic crystals},\
  }\href {https://doi.org/10.1103/PhysRevLett.125.133603} {\bibfield  {journal}
  {\bibinfo  {journal} {Phys. Rev. Lett.}\ }\textbf {\bibinfo {volume} {125}},\
  \bibinfo {pages} {133603} (\bibinfo {year} {2020})}\BibitemShut {NoStop}%
\bibitem [{\citenamefont {Cui}\ \emph {et~al.}(2022)\citenamefont {Cui},
  \citenamefont {Zhang}, \citenamefont {Zhang},\ and\ \citenamefont
  {Chan}}]{Cui36}%
  \BibitemOpen
  \bibfield  {author} {\bibinfo {author} {\bibfnamefont {X.}~\bibnamefont
  {Cui}}, \bibinfo {author} {\bibfnamefont {R.-Y.}\ \bibnamefont {Zhang}},
  \bibinfo {author} {\bibfnamefont {Z.-Q.}\ \bibnamefont {Zhang}},\ and\
  \bibinfo {author} {\bibfnamefont {C.~T.}\ \bibnamefont {Chan}},\ }\bibfield
  {title} {\bibinfo {title} {Photonic ${Z}_{2}$ topological {A}nderson
  insulators},\ }\href {https://doi.org/10.1103/PhysRevLett.129.043902}
  {\bibfield  {journal} {\bibinfo  {journal} {Phys. Rev. Lett.}\ }\textbf
  {\bibinfo {volume} {129}},\ \bibinfo {pages} {043902} (\bibinfo {year}
  {2022})}\BibitemShut {NoStop}%
\bibitem [{\citenamefont {Zhou}\ \emph {et~al.}(2020)\citenamefont {Zhou},
  \citenamefont {Liu}, \citenamefont {Ren}, \citenamefont {Yang}, \citenamefont
  {Xue}, \citenamefont {Bi}, \citenamefont {Deng}, \citenamefont {Chong},\ and\
  \citenamefont {Zhang}}]{Zhou37}%
  \BibitemOpen
  \bibfield  {author} {\bibinfo {author} {\bibfnamefont {P.}~\bibnamefont
  {Zhou}}, \bibinfo {author} {\bibfnamefont {G.-G.}\ \bibnamefont {Liu}},
  \bibinfo {author} {\bibfnamefont {X.}~\bibnamefont {Ren}}, \bibinfo {author}
  {\bibfnamefont {Y.}~\bibnamefont {Yang}}, \bibinfo {author} {\bibfnamefont
  {H.}~\bibnamefont {Xue}}, \bibinfo {author} {\bibfnamefont {L.}~\bibnamefont
  {Bi}}, \bibinfo {author} {\bibfnamefont {L.}~\bibnamefont {Deng}}, \bibinfo
  {author} {\bibfnamefont {Y.}~\bibnamefont {Chong}},\ and\ \bibinfo {author}
  {\bibfnamefont {B.}~\bibnamefont {Zhang}},\ }\bibfield  {title} {\bibinfo
  {title} {Photonic amorphous topological insulator},\ }\href
  {https://doi.org/10.1038/s41377-020-00368-7} {\bibfield  {journal} {\bibinfo
  {journal} {Light Sci. Appl.}\ }\textbf {\bibinfo {volume} {9}},\ \bibinfo
  {pages} {133} (\bibinfo {year} {2020})}\BibitemShut {NoStop}%
\bibitem [{\citenamefont {Yang}\ \emph
  {et~al.}(2019{\natexlab{b}})\citenamefont {Yang}, \citenamefont {Zhang},
  \citenamefont {Wu}, \citenamefont {Dong}, \citenamefont {Yan},\ and\
  \citenamefont {Zhang}}]{Bing38}%
  \BibitemOpen
  \bibfield  {author} {\bibinfo {author} {\bibfnamefont {B.}~\bibnamefont
  {Yang}}, \bibinfo {author} {\bibfnamefont {H.}~\bibnamefont {Zhang}},
  \bibinfo {author} {\bibfnamefont {T.}~\bibnamefont {Wu}}, \bibinfo {author}
  {\bibfnamefont {R.}~\bibnamefont {Dong}}, \bibinfo {author} {\bibfnamefont
  {X.}~\bibnamefont {Yan}},\ and\ \bibinfo {author} {\bibfnamefont
  {X.}~\bibnamefont {Zhang}},\ }\bibfield  {title} {\bibinfo {title}
  {Topological states in amorphous magnetic photonic lattices},\ }\href
  {https://doi.org/10.1103/PhysRevB.99.045307} {\bibfield  {journal} {\bibinfo
  {journal} {Phys. Rev. B}\ }\textbf {\bibinfo {volume} {99}},\ \bibinfo
  {pages} {045307} (\bibinfo {year} {2019}{\natexlab{b}})}\BibitemShut
  {NoStop}%
\bibitem [{\citenamefont {Li}\ \emph {et~al.}(2009)\citenamefont {Li},
  \citenamefont {Chu}, \citenamefont {Jain},\ and\ \citenamefont
  {Shen}}]{Li39}%
  \BibitemOpen
  \bibfield  {author} {\bibinfo {author} {\bibfnamefont {J.}~\bibnamefont
  {Li}}, \bibinfo {author} {\bibfnamefont {R.-L.}\ \bibnamefont {Chu}},
  \bibinfo {author} {\bibfnamefont {J.~K.}\ \bibnamefont {Jain}},\ and\
  \bibinfo {author} {\bibfnamefont {S.-Q.}\ \bibnamefont {Shen}},\ }\bibfield
  {title} {\bibinfo {title} {Topological {A}nderson insulator},\ }\href
  {https://doi.org/10.1103/PhysRevLett.102.136806} {\bibfield  {journal}
  {\bibinfo  {journal} {Phys. Rev. Lett.}\ }\textbf {\bibinfo {volume} {102}},\
  \bibinfo {pages} {136806} (\bibinfo {year} {2009})}\BibitemShut {NoStop}%
\bibitem [{\citenamefont {Groth}\ \emph {et~al.}(2009)\citenamefont {Groth},
  \citenamefont {Wimmer}, \citenamefont {Akhmerov}, \citenamefont
  {Tworzyd\l{}o},\ and\ \citenamefont {Beenakker}}]{Groth40}%
  \BibitemOpen
  \bibfield  {author} {\bibinfo {author} {\bibfnamefont {C.~W.}\ \bibnamefont
  {Groth}}, \bibinfo {author} {\bibfnamefont {M.}~\bibnamefont {Wimmer}},
  \bibinfo {author} {\bibfnamefont {A.~R.}\ \bibnamefont {Akhmerov}}, \bibinfo
  {author} {\bibfnamefont {J.}~\bibnamefont {Tworzyd\l{}o}},\ and\ \bibinfo
  {author} {\bibfnamefont {C.~W.~J.}\ \bibnamefont {Beenakker}},\ }\bibfield
  {title} {\bibinfo {title} {Theory of the topological {A}nderson insulator},\
  }\href {https://doi.org/10.1103/PhysRevLett.103.196805} {\bibfield  {journal}
  {\bibinfo  {journal} {Phys. Rev. Lett.}\ }\textbf {\bibinfo {volume} {103}},\
  \bibinfo {pages} {196805} (\bibinfo {year} {2009})}\BibitemShut {NoStop}%
\bibitem [{\citenamefont {Meier}\ \emph {et~al.}(2018)\citenamefont {Meier},
  \citenamefont {An}, \citenamefont {Dauphin}, \citenamefont {Maffei},
  \citenamefont {Massignan}, \citenamefont {Hughes},\ and\ \citenamefont
  {Gadway}}]{Meier41}%
  \BibitemOpen
  \bibfield  {author} {\bibinfo {author} {\bibfnamefont {E.~J.}\ \bibnamefont
  {Meier}}, \bibinfo {author} {\bibfnamefont {F.~A.}\ \bibnamefont {An}},
  \bibinfo {author} {\bibfnamefont {A.}~\bibnamefont {Dauphin}}, \bibinfo
  {author} {\bibfnamefont {M.}~\bibnamefont {Maffei}}, \bibinfo {author}
  {\bibfnamefont {P.}~\bibnamefont {Massignan}}, \bibinfo {author}
  {\bibfnamefont {T.~L.}\ \bibnamefont {Hughes}},\ and\ \bibinfo {author}
  {\bibfnamefont {B.}~\bibnamefont {Gadway}},\ }\bibfield  {title} {\bibinfo
  {title} {Observation of the topological {A}nderson insulator in disordered
  atomic wires},\ }\href {https://doi.org/10.1126/science.aat3406} {\bibfield
  {journal} {\bibinfo  {journal} {Science}\ }\textbf {\bibinfo {volume}
  {362}},\ \bibinfo {pages} {929} (\bibinfo {year} {2018})}\BibitemShut
  {NoStop}%
\bibitem [{\citenamefont {Mott}\ and\ \citenamefont {Jones}(1958)}]{Mott42}%
  \BibitemOpen
  \bibfield  {author} {\bibinfo {author} {\bibfnamefont {N.~F.}\ \bibnamefont
  {Mott}}\ and\ \bibinfo {author} {\bibfnamefont {H.}~\bibnamefont {Jones}},\
  }\href@noop {} {\emph {\bibinfo {title} {The theory of the properties of
  metals and alloys}}}\ (\bibinfo  {publisher} {Courier Dover Publications},\
  \bibinfo {year} {1958})\BibitemShut {NoStop}%
\bibitem [{\citenamefont {George}\ \emph {et~al.}(2019)\citenamefont {George},
  \citenamefont {Raabe},\ and\ \citenamefont {Ritchie}}]{George43}%
  \BibitemOpen
  \bibfield  {author} {\bibinfo {author} {\bibfnamefont {E.~P.}\ \bibnamefont
  {George}}, \bibinfo {author} {\bibfnamefont {D.}~\bibnamefont {Raabe}},\ and\
  \bibinfo {author} {\bibfnamefont {R.~O.}\ \bibnamefont {Ritchie}},\
  }\bibfield  {title} {\bibinfo {title} {High-entropy alloys},\ }\href
  {https://doi.org/10.1038/s41578-019-0121-4} {\bibfield  {journal} {\bibinfo
  {journal} {Nat. Rev. Mater.}\ }\textbf {\bibinfo {volume} {4}},\ \bibinfo
  {pages} {515} (\bibinfo {year} {2019})}\BibitemShut {NoStop}%
\bibitem [{\citenamefont {Hart}\ \emph {et~al.}(2021)\citenamefont {Hart},
  \citenamefont {Mueller}, \citenamefont {Toher},\ and\ \citenamefont
  {Curtarolo}}]{Hart44}%
  \BibitemOpen
  \bibfield  {author} {\bibinfo {author} {\bibfnamefont {G.~L.~W.}\
  \bibnamefont {Hart}}, \bibinfo {author} {\bibfnamefont {T.}~\bibnamefont
  {Mueller}}, \bibinfo {author} {\bibfnamefont {C.}~\bibnamefont {Toher}},\
  and\ \bibinfo {author} {\bibfnamefont {S.}~\bibnamefont {Curtarolo}},\
  }\bibfield  {title} {\bibinfo {title} {Machine learning for alloys},\ }\href
  {https://doi.org/10.1038/s41578-021-00340-w} {\bibfield  {journal} {\bibinfo
  {journal} {Nat. Rev. Mater.}\ }\textbf {\bibinfo {volume} {6}},\ \bibinfo
  {pages} {730} (\bibinfo {year} {2021})}\BibitemShut {NoStop}%
\bibitem [{\citenamefont {Skirlo}\ \emph {et~al.}(2014)\citenamefont {Skirlo},
  \citenamefont {Lu},\ and\ \citenamefont {Solja\ifmmode \check{c}\else
  \v{c}\fi{}i\ifmmode~\acute{c}\else \'{c}\fi{}}}]{Skirlo45}%
  \BibitemOpen
  \bibfield  {author} {\bibinfo {author} {\bibfnamefont {S.~A.}\ \bibnamefont
  {Skirlo}}, \bibinfo {author} {\bibfnamefont {L.}~\bibnamefont {Lu}},\ and\
  \bibinfo {author} {\bibfnamefont {M.}~\bibnamefont {Solja\ifmmode
  \check{c}\else \v{c}\fi{}i\ifmmode~\acute{c}\else \'{c}\fi{}}},\ }\bibfield
  {title} {\bibinfo {title} {Multimode one-way waveguides of large chern
  numbers},\ }\href {https://doi.org/10.1103/PhysRevLett.113.113904} {\bibfield
   {journal} {\bibinfo  {journal} {Phys. Rev. Lett.}\ }\textbf {\bibinfo
  {volume} {113}},\ \bibinfo {pages} {113904} (\bibinfo {year}
  {2014})}\BibitemShut {NoStop}%
\bibitem [{\citenamefont {Skirlo}\ \emph {et~al.}(2015)\citenamefont {Skirlo},
  \citenamefont {Lu}, \citenamefont {Igarashi}, \citenamefont {Yan},
  \citenamefont {Joannopoulos},\ and\ \citenamefont {Solja\ifmmode
  \check{c}\else \v{c}\fi{}i\ifmmode~\acute{c}\else \'{c}\fi{}}}]{Skirlo46}%
  \BibitemOpen
  \bibfield  {author} {\bibinfo {author} {\bibfnamefont {S.~A.}\ \bibnamefont
  {Skirlo}}, \bibinfo {author} {\bibfnamefont {L.}~\bibnamefont {Lu}}, \bibinfo
  {author} {\bibfnamefont {Y.}~\bibnamefont {Igarashi}}, \bibinfo {author}
  {\bibfnamefont {Q.}~\bibnamefont {Yan}}, \bibinfo {author} {\bibfnamefont
  {J.}~\bibnamefont {Joannopoulos}},\ and\ \bibinfo {author} {\bibfnamefont
  {M.}~\bibnamefont {Solja\ifmmode \check{c}\else
  \v{c}\fi{}i\ifmmode~\acute{c}\else \'{c}\fi{}}},\ }\bibfield  {title}
  {\bibinfo {title} {Experimental observation of large chern numbers in
  photonic crystals},\ }\href {https://doi.org/10.1103/PhysRevLett.115.253901}
  {\bibfield  {journal} {\bibinfo  {journal} {Phys. Rev. Lett.}\ }\textbf
  {\bibinfo {volume} {115}},\ \bibinfo {pages} {253901} (\bibinfo {year}
  {2015})}\BibitemShut {NoStop}%
\bibitem [{sup()}]{supp}%
  \BibitemOpen
  \href@noop {} {}\bibinfo {howpublished} {See Supplemental Material at
  {http://link.aps.org/supplemental/XXX}, for (1) methods, (2) randomly
  selected configurations of substitutional photonic alloys, (3) simulated
  field distribution in substitutional photonic alloys ${\rm A}_{1-x}{\rm B}_x$
  and ${\rm Air}_{1-x}{\rm B}_x$, (4) experimentally measured edge transmission
  in substitutional photonic alloys with varying magnetic rod concentrations,
  (5) simulated density of states (DOS) gap in substitutional photonic alloys,
  (6) measurement results of edge field distribution in substitutional photonic
  alloys with varying magnetic rod concentrations, (7) scattering approach for
  classifying topological invariants in photonic alloys, (8) upper and lower
  phase boundaries of the topological gap in substitutional photonic alloys,
  (9) generalized Haldane model, (10) interstitial photonic alloys, {(11) band
  structures of the honeycomb and Kagome lattices,} which includes Refs.
  \cite{Haldane48,Datta49}.}\BibitemShut {Stop}%
\bibitem [{\citenamefont {Haldane}(1988)}]{Haldane48}%
  \BibitemOpen
  \bibfield  {author} {\bibinfo {author} {\bibfnamefont {F.~D.~M.}\
  \bibnamefont {Haldane}},\ }\bibfield  {title} {\bibinfo {title} {Model for a
  quantum {H}all effect without {L}andau levels: Condensed-matter realization
  of the ``parity anomal''},\ }\href
  {https://doi.org/10.1103/PhysRevLett.61.2015} {\bibfield  {journal} {\bibinfo
   {journal} {Phys. Rev. Lett.}\ }\textbf {\bibinfo {volume} {61}},\ \bibinfo
  {pages} {2015} (\bibinfo {year} {1988})}\BibitemShut {NoStop}%
\bibitem [{\citenamefont {Datta}(1997)}]{Datta49}%
  \BibitemOpen
  \bibfield  {author} {\bibinfo {author} {\bibfnamefont {S.}~\bibnamefont
  {Datta}},\ }\href@noop {} {\emph {\bibinfo {title} {Electronic transport in
  mesoscopic systems}}}\ (\bibinfo  {publisher} {Cambridge University Press},\
  \bibinfo {year} {1997})\BibitemShut {NoStop}%
\end{thebibliography}%

\end{document}